\documentclass{article}

\usepackage{arxiv}

\usepackage[utf8]{inputenc} % allow utf-8 input
\usepackage[T1]{fontenc}    % use 8-bit T1 fonts
\usepackage{hyperref}       % hyperlinks
\usepackage{url}            % simple URL typesetting
\usepackage{booktabs}       % professional-quality tables
\usepackage{amsfonts}       % blackboard math symbols
\usepackage{nicefrac}       % compact symbols for 1/2, etc.
\usepackage{microtype}      % microtypography
\usepackage{lipsum}
\usepackage{graphicx}

\usepackage{authblk}

\title{DeepReg: a deep learning toolkit \\ for medical image registration}

\author[1,2,3]{Yunguan Fu}
\author[1,2]{Nina Montaña Brown}
\author[1,2]{Shaheer U. Saeed}
\author[2]{Adrià Casamitjana}
\author[1,2]{Zachary M. C. Baum}
\author[1,4]{\authorcr Rémi Delaunay}
\author[1,2]{Qianye Yang}
\author[1,2]{Alexander Grimwood}
\author[1]{Zhe Min}
\author[2]{Stefano B. Blumberg}
\author[2,5,6]{\authorcr Juan Eugenio Iglesias}
\author[1,2]{Dean C. Barratt}
\author[1,2]{Ester Bonmati}
\author[2]{\authorcr Daniel C. Alexander}
\author[1,2]{Matthew J. Clarkson}
\author[4]{Tom Vercauteren}
\author[1,2]{Yipeng Hu}

\affil[1]{Wellcome/EPSRC Centre for Surgical and Interventional Sciences, University College London, London, UK}
\affil[2]{Centre for Medical Image Computing, University College London, London, UK}
\affil[3]{InstaDeep, London, UK}
\affil[4]{Department of Surgical \& Interventional Engineering, King’s College London, London, UK}
\affil[5]{Martinos Center for Biomedical Imaging, Massachusetts General Hospital and Harvard Medical School, Boston, USA}
\affil[6]{Computer Science and Artificial Intelligence Laboratory, Massachusetts Institute of Technology, Boston, USA}

\date{}

\renewcommand{\headeright}{}
\renewcommand{\undertitle}{}
\renewcommand{\shorttitle}{DeepReg: a deep learning toolkit for medical image registration}

\hypersetup{
pdftitle={DeepReg: a deep learning toolkit for medical image registration},
}

\begin{document}

\maketitle

\begin{abstract}
DeepReg (\url{https://github.com/DeepRegNet/DeepReg}) is a community-supported open-source toolkit for research and education in medical image registration using deep learning.
\end{abstract}

\section{Summary}
Image fusion is a fundamental task in medical image analysis and computer-assisted intervention. Medical image registration, computational algorithms that align different images together \cite{hill2001medical}, has in recent years turned the research attention towards deep learning. Indeed, the representation ability to learn from population data with deep neural networks has opened new possibilities for improving registration generalisability by mitigating difficulties in designing hand-engineered image features and similarity measures for many real-world clinical applications \cite{haskins2020deep,fu2020deep}. In addition, its fast inference can substantially accelerate registration execution for time-critical tasks.

DeepReg is a Python package using TensorFlow \cite{tensorflow2015-whitepaper} that implements multiple registration algorithms and a set of predefined dataset loaders, supporting both labelled- and unlabelled data. DeepReg also provides command-line tool options that enable basic and advanced functionalities for model training, prediction and image warping. These implementations, together with their documentation, tutorials and demos, aim to simplify workflows for prototyping and developing novel methodology, utilising latest development and accessing quality research advances. DeepReg is unit tested and a set of customised contributor guidelines are provided to facilitate community contributions.

A submission to the MICCAI Educational Challenge has utilised the DeepReg code and demos to explore the link between classical algorithms and deep-learning-based methods \cite{brown2020introduction}, while a recently published research work investigated temporal changes in prostate cancer imaging, by using a longitudinal registration adapted from the DeepReg code \cite{yang2020longitudinal}.

\section{Statement of need}
Currently, popular packages focusing on deep learning methods for medical imaging, such as NiftyNet \cite{gibson2018niftynet} and MONAI (\url{https://monai.io/}), do not support image registration. The existing open-sourced registration projects either implement specific published algorithms without automated testing, such as the VoxelMorph \cite{balakrishnan2019voxelmorph}, or focus on classical methods, such as NiftiReg \cite{modat2010fast}, SimpleElastix \cite{marstal2016simpleelastix} and AirLab \cite{sandkuhler2018airlab}. Therefore an open-sourced project focusing on image registration with deep learning is much needed for general research and education purposes.

\section{Implementation}
DeepReg implements a framework for unsupervised learning \cite{de2019deep,balakrishnan2019voxelmorph}, weakly-supervised learning \cite{hu2018label,hu2018weakly} and their combinations and variants, e.g. \cite{hu2019conditional}. Many options are included for major components of these approaches, such as different image- and label dissimilarity functions, transformation models \cite{ashburner2007fast,vercauteren2009diffeomorphic,hill2001medical}, deformation regularisation \cite{rueckert1999nonrigid} and different neural network architectures \cite{hu2018weakly,he2016deep,simonyan2014very}. Details of the implemented methods are described in the documentation. The provided dataset loaders adopt staged random sampling strategy to ensure unbiased learning from groups, images and labels \cite{hu2018weakly,yang2020longitudinal}. These algorithmic components together with the flexible dataset loaders are building blocks of many other registration tasks, such as group-wise registration and morphological template construction \cite{dalca2019learning,siebert2020deep,luo2020mvmm}.

\section{DeepReg Demos}
In addition to the tutorials and documentation, DeepReg provides a collection of demonstrations, DeepReg Demos, using open-accessible data with real-world clinical applications.

\subsection{Paired images}
Many clinical applications for tracking organ motion and other temporal changes require intra-subject single-modality image registration. Registering lung CT images for the same patient, acquired at expiratory and inspiratory phases \cite{hering_alessa_2020_3835682}, is such an example of both unsupervised (without labels) and combined supervision (trained with additional label dissimilarity based on anatomical segmentation). Furthermore, registering prostate MR, acquired before surgery, and intra-operative ultrasound images is an example of weakly-supervised learning for multimodal image registration \cite{hu2018weakly}. Another DeepReg Demo illustrates MR-to-ultrasound image registration is to track tissue deformation and brain tumour resection during neurosurgery \cite{xiao2017resect}.

\subsection{Unpaired images}
Unpaired images are found in applications such as single-modality inter-subject registration. One demo registers different brain MR images from different subjects \cite{simpson2019large}, fundamental to population studies. Two other applications align unpaired inter-subject CT images for lung \cite{hering_alessa_2020_3835682} and abdominal organs \cite{adrian_dalca_2020_3715652}. Additionally, the support for cross-validation in DeepReg has been included in a demo, which registers 3D ultrasound images from different prostate cancer patients.

\subsection{Grouped images}
Unpaired images may also be grouped in applications such as single-modality intra-subject registration. In this case, each subject has multiple images acquired, for instance, at two or more time points. For demonstration, multi-sequence cardiac MR images, acquired from myocardial infarction patients \cite{zhuang2020cardiac}, are registered, where multiple images within each subject are considered as grouped images. Prostate longitudinal MR registration is proposed to track the cancer progression during active surveillance programme \cite{yang2020longitudinal}. Using segmentation from this application, another demo application illustrates aligning intra-patient prostate gland masks - also an example of feature-based registration based on deep learning.

\section{Conclusion}
DeepReg provides a collection of deep learning algorithms and dataset loaders to train image registration networks, which provides a reference of basic functionalities. In its permissible open-source format, DeepReg not only provides a tool for scientific research and higher education, but also welcomes contributions from wider communities.

\section*{Acknowledgements}

This work is supported by the Wellcome/EPSRC Centre for Interventional and Surgical Sciences (203145Z/16/Z). Support was also from the Engineering and Physical Sciences Research Council (EPSRC) (EP/M020533/1, NS/A000049/1), National Institute for Health Research University College London Hospitals Biomedical Research Centre and Wellcome Trust (203148/Z/16/Z). TV is supported by a Medtronic / Royal Academy of Engineering Research Chair (RCSRF1819\textbackslash7\textbackslash34). NMB, ZB, RD are also supported by the EPSRC CDT i4health (EP/S021930/1). ZB is supported by the Natural Sciences and Engineering Research Council of Canada Postgraduate Scholarships-Doctoral Program and the University College London Overseas and Graduate Research Scholarships.

\bibliographystyle{unsrt}  
\bibliography{references}

\begin{thebibliography}{10}

\bibitem{hill2001medical}
Derek~LG Hill, Philipp~G Batchelor, Mark Holden, and David~J Hawkes.
\newblock Medical image registration.
\newblock {\em Physics in medicine \& biology}, 46(3):R1, 2001.

\bibitem{haskins2020deep}
Grant Haskins, Uwe Kruger, and Pingkun Yan.
\newblock Deep learning in medical image registration: a survey.
\newblock {\em Machine Vision and Applications}, 31(1):8, 2020.

\bibitem{fu2020deep}
Yabo Fu, Yang Lei, Tonghe Wang, Walter~J Curran, Tian Liu, and Xiaofeng Yang.
\newblock Deep learning in medical image registration: a review.
\newblock {\em Physics in Medicine \& Biology}, 2020.

\bibitem{tensorflow2015-whitepaper}
Mart\'{\i}n Abadi, Ashish Agarwal, Paul Barham, Eugene Brevdo, Zhifeng Chen,
  Craig Citro, Greg~S. Corrado, Andy Davis, Jeffrey Dean, Matthieu Devin,
  Sanjay Ghemawat, Ian Goodfellow, Andrew Harp, Geoffrey Irving, Michael Isard,
  Yangqing Jia, Rafal Jozefowicz, Lukasz Kaiser, Manjunath Kudlur, Josh
  Levenberg, Dandelion Man\'{e}, Rajat Monga, Sherry Moore, Derek Murray, Chris
  Olah, Mike Schuster, Jonathon Shlens, Benoit Steiner, Ilya Sutskever, Kunal
  Talwar, Paul Tucker, Vincent Vanhoucke, Vijay Vasudevan, Fernanda Vi\'{e}gas,
  Oriol Vinyals, Pete Warden, Martin Wattenberg, Martin Wicke, Yuan Yu, and
  Xiaoqiang Zheng.
\newblock {TensorFlow}: Large-scale machine learning on heterogeneous systems,
  2015.
\newblock Software available from tensorflow.org.

\bibitem{brown2020introduction}
Nina Montana~Brown, Yunguan Fu, Shaheer~U. Saeed, Adrià Casamitjana, Zachary
  M.~C. Baum, Rémi Delaunay, Qianye Yang, Alexander Grimwood, Zhe Min, Ester
  Bonmati, Tom Vercauteren, Matthew~J. Clarkson, and Yipeng Hu.
\newblock Introduction to medical image registration with deepreg, between old
  and new.
\newblock 2020.

\bibitem{yang2020longitudinal}
Qianye Yang, Yunguan Fu, Francesco Giganti, Nooshin Ghavami, Qingchao Chen,
  J.~Alison Noble, Tom Vercauteren, Dean Barratt, and Yipeng Hu.
\newblock Longitudinal image registration with temporal-order and
  subject-specificity discrimination, 2020.

\bibitem{gibson2018niftynet}
Eli Gibson, Wenqi Li, Carole Sudre, Lucas Fidon, Dzhoshkun~I Shakir, Guotai
  Wang, Zach Eaton-Rosen, Robert Gray, Tom Doel, Yipeng Hu, et~al.
\newblock Niftynet: a deep-learning platform for medical imaging.
\newblock {\em Computer methods and programs in biomedicine}, 158:113--122,
  2018.

\bibitem{balakrishnan2019voxelmorph}
Guha Balakrishnan, Amy Zhao, Mert~R Sabuncu, John Guttag, and Adrian~V Dalca.
\newblock Voxelmorph: a learning framework for deformable medical image
  registration.
\newblock {\em IEEE transactions on medical imaging}, 38(8):1788--1800, 2019.

\bibitem{modat2010fast}
Marc Modat, Gerard~R Ridgway, Zeike~A Taylor, Manja Lehmann, Josephine Barnes,
  David~J Hawkes, Nick~C Fox, and S{\'e}bastien Ourselin.
\newblock Fast free-form deformation using graphics processing units.
\newblock {\em Computer methods and programs in biomedicine}, 98(3):278--284,
  2010.

\bibitem{marstal2016simpleelastix}
Kasper Marstal, Floris Berendsen, Marius Staring, and Stefan Klein.
\newblock Simpleelastix: A user-friendly, multi-lingual library for medical
  image registration.
\newblock pages 134--142, 2016.

\bibitem{sandkuhler2018airlab}
Robin Sandk{\"u}hler, Christoph Jud, Simon Andermatt, and Philippe~C Cattin.
\newblock Airlab: autograd image registration laboratory.
\newblock {\em arXiv preprint arXiv:1806.09907}, 2018.

\bibitem{de2019deep}
Bob~D de~Vos, Floris~F Berendsen, Max~A Viergever, Hessam Sokooti, Marius
  Staring, and Ivana I{\v{s}}gum.
\newblock A deep learning framework for unsupervised affine and deformable
  image registration.
\newblock {\em Medical image analysis}, 52:128--143, 2019.

\bibitem{hu2018label}
Yipeng Hu, Marc Modat, Eli Gibson, Nooshin Ghavami, Ester Bonmati, Caroline~M
  Moore, Mark Emberton, J~Alison Noble, Dean~C Barratt, and Tom Vercauteren.
\newblock Label-driven weakly-supervised learning for multimodal deformable
  image registration.
\newblock In {\em 2018 IEEE 15th International Symposium on Biomedical Imaging
  (ISBI 2018)}, pages 1070--1074. IEEE, 2018.

\bibitem{hu2018weakly}
Yipeng Hu, Marc Modat, Eli Gibson, Wenqi Li, Nooshin Ghavami, Ester Bonmati,
  Guotai Wang, Steven Bandula, Caroline~M Moore, Mark Emberton, et~al.
\newblock Weakly-supervised convolutional neural networks for multimodal image
  registration.
\newblock {\em Medical image analysis}, 49:1--13, 2018.

\bibitem{hu2019conditional}
Yipeng Hu, Eli Gibson, Dean~C Barratt, Mark Emberton, J~Alison Noble, and Tom
  Vercauteren.
\newblock Conditional segmentation in lieu of image registration.
\newblock In {\em International Conference on Medical Image Computing and
  Computer-Assisted Intervention}, pages 401--409. Springer, 2019.

\bibitem{ashburner2007fast}
John Ashburner.
\newblock A fast diffeomorphic image registration algorithm.
\newblock {\em Neuroimage}, 38(1):95--113, 2007.

\bibitem{vercauteren2009diffeomorphic}
Tom Vercauteren, Xavier Pennec, Aymeric Perchant, and Nicholas Ayache.
\newblock Diffeomorphic demons: Efficient non-parametric image registration.
\newblock {\em NeuroImage}, 45(1):S61--S72, 2009.

\bibitem{rueckert1999nonrigid}
Daniel Rueckert, Luke~I Sonoda, Carmel Hayes, Derek~LG Hill, Martin~O Leach,
  and David~J Hawkes.
\newblock Nonrigid registration using free-form deformations: application to
  breast mr images.
\newblock {\em IEEE transactions on medical imaging}, 18(8):712--721, 1999.

\bibitem{he2016deep}
Kaiming He, Xiangyu Zhang, Shaoqing Ren, and Jian Sun.
\newblock Deep residual learning for image recognition.
\newblock In {\em Proceedings of the IEEE conference on computer vision and
  pattern recognition}, pages 770--778, 2016.

\bibitem{simonyan2014very}
Karen Simonyan and Andrew Zisserman.
\newblock Very deep convolutional networks for large-scale image recognition.
\newblock {\em arXiv preprint arXiv:1409.1556}, 2014.

\bibitem{dalca2019learning}
Adrian Dalca, Marianne Rakic, John Guttag, and Mert Sabuncu.
\newblock Learning conditional deformable templates with convolutional
  networks.
\newblock In {\em Advances in neural information processing systems}, pages
  806--818, 2019.

\bibitem{siebert2020deep}
Hanna Siebert and Mattias~P Heinrich.
\newblock Deep groupwise registration of mri using deforming autoencoders.
\newblock In {\em Bildverarbeitung f{\"u}r die Medizin 2020}, pages 236--241.
  Springer, 2020.

\bibitem{luo2020mvmm}
Xinzhe Luo and Xiahai Zhuang.
\newblock Mvmm-regnet: A new image registration framework based on multivariate
  mixture model and neural network estimation.
\newblock {\em arXiv preprint arXiv:2006.15573}, 2020.

\bibitem{hering_alessa_2020_3835682}
Alessa Hering, Keelin Murphy, and Bram van Ginneken.
\newblock {Lean2Reg Challenge: CT Lung Registration - Training Data}, May 2020.

\bibitem{xiao2017resect}
Yiming Xiao, Maryse Fortin, Geirmund Unsg{\aa}rd, Hassan Rivaz, and Ingerid
  Reinertsen.
\newblock Retrospective evaluation of cerebral tumors (resect): A clinical
  database of pre-operative mri and intra-operative ultrasound in low-grade
  glioma surgeries.
\newblock {\em Medical physics}, 44(7):3875--3882, 2017.

\bibitem{simpson2019large}
Amber~L Simpson, Michela Antonelli, Spyridon Bakas, Michel Bilello, Keyvan
  Farahani, Bram Van~Ginneken, Annette Kopp-Schneider, Bennett~A Landman, Geert
  Litjens, Bjoern Menze, et~al.
\newblock A large annotated medical image dataset for the development and
  evaluation of segmentation algorithms.
\newblock {\em arXiv preprint arXiv:1902.09063}, 2019.

\bibitem{adrian_dalca_2020_3715652}
Adrian Dalca, Yipeng Hu, Tom Vercauteren, Mattias Heinrich, Lasse Hansen, Marc
  Modat, Bob de~Vos, Yiming Xiao, Hassan Rivaz, Matthieu Chabanas, Ingerid
  Reinertsen, Bennett Landman, Jorge Cardoso, Bram van Ginneken, Alessa Hering,
  and Keelin Murphy.
\newblock Learn2reg - the challenge, March 2020.

\bibitem{zhuang2020cardiac}
Xiahai Zhuang, Jiahang Xu, Xinzhe Luo, Chen Chen, Cheng Ouyang, Daniel
  Rueckert, Victor~M Campello, Karim Lekadir, Sulaiman Vesal, Nishant
  RaviKumar, et~al.
\newblock Cardiac segmentation on late gadolinium enhancement mri: A benchmark
  study from multi-sequence cardiac mr segmentation challenge.
\newblock {\em arXiv preprint arXiv:2006.12434}, 2020.

\end{thebibliography}

\end{document}